\documentclass[pre, floatfix, nofootinbib, letterpaper, preprint]{revtex4}
\usepackage{float}
\usepackage[caption=false]{subfig}
\usepackage{amssymb, amsmath}
\usepackage[colorlinks,citecolor=blue,urlcolor=blue]{hyperref}
\usepackage{color}
\usepackage{graphicx}
\newcommand{\InsertFig}[4]
{\begin{figure}[h!t]
         \includegraphics[width=#4\columnwidth]{./#1}
       \caption{{\footnotesize  #2}
       \label{fig:#3}}
\end{figure}}

\newcommand{\InsertFigTwo}[5] {
\begin{figure*}[htb]
       \centerline{
         \includegraphics[width=#5\columnwidth]{./#1}
         \hskip 0.5in
         \includegraphics[width=#5\columnwidth]{./#2}
       }
       \caption{{\footnotesize  #3}
       \label{fig:#4}}
\end{figure*}}

\newcommand{\Eq}[1]{(\ref{eq:#1})}
\newcommand{\Sec}[1]{\S \ref{sec:#1}}
\newcommand{\Fig}[1]{Fig.~\ref{fig:#1}}

\begin{document}

\title{Diffusion For Ensembles of Standard Maps.}
\author{Or Alus}
\email{oralus@tx.technion.ac.il}
\author{Shmuel Fishman}
\email{fishman@physics.technion.ac.il}
\affiliation{Physics Department\\ Technion--Israel Institute of Technology\\ Haifa 32000, Israel}

\date{\today}

\begin{abstract}
	Two types of random evolution processes are studied for ensembles of the standard map with driving parameter $K$ that determines its degree of stochasticity. For one type
of processes the parameter $K$ is chosen at random from a Gaussian
distribution and is then kept fixed, while for the other type it varies
from step to step. In addition, noise that can be arbitrarily weak is added. The ensemble average and the average over noise of the diffusion coefficient is calculated for both types of processes. These two types of processes are relevant for two types of experimental situations as explained in the paper. Both types of processes destroy fine details of the dynamics, and the second process is found to be more effective in destroying the fine details. We hope that this work is a step in the efforts for developing a statistical theory for systems with mixed phase space (regular in some parts and chaotic in other parts).
\end{abstract}

\pacs{XXXXXX}
\vspace*{1ex}
\noindent

\maketitle

\section{Introduction}\label{sec:intro}
Typically physical systems are modeled by Hamiltonians or maps leading to dynamics in mixed phase space \cite{Meiss1992,Tabor1989,lichtenberg1983}. In such a phase space, the dynamics in some parts is chaotic and in other parts it is regular. Here we consider the case of conservative dynamics (area in phase space is conserved). The phase space exhibits structures on all scales. These fine details are typically very sensitive to the values of parameters, far beyond experimental resolution.  In mixed systems transport is affected by ``sticking'' to regular structures such as islands chains, see also \cite{Zaslavsky1997,Zaslavsky2000}. By sticking we mean that a trajectory is trapped for a very long time near some structure. The regular structures are typically surrounded by broken invariant circles called ``cantori'' which function as barriers to the flux of chaotic trajectories.   Chirikov and Shepelyansky \cite{Physics1984} studied the decay of correlations near the critical point $K_c$ of the standard map, where chaos becomes unbounded (see \Eq{Standard map} and discussion that follows). They observed an algebraic decay of correlations for long times. This calls for a statistical description of such systems \cite{Alus}.

Although time correlations in a specific region of phase space may decay algebraically, rapid decay of time correlations between different parts of phase space should be considered. Therefore a statistical approach may be applicable to a single system. A comprehensive model for transport in such systems was proposed by Meiss and Ott \cite{Meiss1985}, where a construction of a distribution of fluxes through different structures was introduced. In this way a complicated deterministic process was replaced by relatively simple random one. The distribution of flux ratios relevant for this process was calculated for the H\'enon Map in Ref. \cite{Alus}. Recently an important contribution was made by Cristadoro and Ketzmerick who demonstrated the universality of the decay of correlations in the framework of the model of \cite{Meiss1985}. They examined an ensemble of such systems by using an arbitrary distribution of transition probabilities in phase space \cite{Cristadoro2008} . Guided by similar ideas, Ceder and Agam \cite{Ceder2013} used diagrammatic methods to calculate the exponent of the decay of correlations and its fluctuations and found that the fluctuations are large. A summary of the exponents of the decay of correlations and relations to exponents characterizing the spreading is presented by Venegeroles \cite{Venegeroles2009}. 

Another approach to treat statistically this effect in such systems without modeling phase space was suggested in the pioneering work of Rechester, Rosenbluth and White (RRW) \cite{Rechester1981,Rechester1980} where noise was used. This enables to define and calculate the diffusion coefficient in phase space. In particular the limit of vanishing noise was found to be meaningful. In these calculations the main effect of noise is to suppress long time ``sticking'' and to enable diffusion. Sometimes it is referred to as regularization of the diffusion process. On the other hand it was found that noise may enhance trapping \cite{PhysRevLett.105.244102}. This motivates calculation of various scenarios and classification of ``noisy" phenomena. A characterization of this type was introduced by Romeiras, Grebogi and Ott \cite{PhysRevA.41.784} (see also \cite{PhysRevE.87.042902}):

Problem 1 (noisy map): For an ensemble of trajectories each encounters different random perturbations.

Problem 2 (random map): For an ensemble of trajectories all encounter the same random perturbation.

In the present work we apply noise like in problem 1, of small variance, and in some cases we consider a random map like in problem 2. The processes we study can be classified into two types:

\textit{Type I}: An ensemble of systems each with a different parameter value that is constant in time.

\textit{Type II}: An ensemble of systems where the parameters vary randomly with time, where each member is like problem 2 (random map).

Randomness of \textit{type I} is relevant for ensembles of devices such as driven Josephson junctions and squids \cite{PhysRevX.3.041029}. Randomness of \textit{type II} may be relevant for atomic billiards \cite{PhysRevLett.86.1518,PhysRevLett.86.1514} where in spite of the experimental efforts the walls of the billiard move during the experiment. 

The noise introduced by RRW (problem 1) results most naturally in experiments, as result of the interaction with the environment. The purpose of the present paper is to explore the statistical effects of the randomness of \textit{type I} and \textit{type II} in the presence of weak noise of the type introduced by RRW. We will explore in particular the question of the effect of the two types of randomness on the fine details of the system.
The noise was introduced by RRW to regularize the map and to be able to define the diffusion coefficient. The reason the result is meaningful, also in the absence of noise is the fact that for short times correlations fall off exponentially, as in the case of idealized fully chaotic systems in the asymptotic infinite time limit \cite{Khodas2000a,Khodas2000b,Venegeroles2008}. This leads to a result similar to the one of RRW \cite{PhysRevA.23.2744}. The main goal of introducing randomness in the present work is to study ensembles of mixed systems rather than achieving regularization for a specific system.   
The calculations in the present paper are performed for the standard map \cite{Chirikov1979a,Tabor1989,lichtenberg1983} .
This map is given in terms of the variables $\theta$ and $J$:
\begin{equation}\label{eq:Standard map}
\begin{array}{c}
\theta_{t+1}=\theta_{t}+J_{t+1}\\
J_{t+1}=J_{t}-K\sin\left(\theta_{t}\right)
\end{array}
\end{equation}
where the parameter $K$ controls the level of chaos. For
$K>K_{c}=0.971635...$ it exhibits diffusion like dynamics in momentum $J$
for most values of $K$. The diffusion coefficient is 
\begin{equation}\label{eq:Diffusion}
D(K)=\lim_{t\to\infty}\tfrac{1}{2t}\left<{\left(J_{t}-J_{0}\right)}^2\right>
\end{equation}
where $\left< \right>$ denotes the average over initial conditions in the chaotic component. If RRW's type of noise is added also averaging over the noise is understood. Fine details of the system are the accelerator modes, that is, for $K$ in the vicinity of $K=2\pi n$,
where $n$ is an integer, acceleration is found for some initial conditions. Effects of noise (problem 1) were studied specifically for accelerator modes \cite{Karney1982425}.
Deviations from diffusion as a result of sticking were studied by Zaslavsky
and Edelman \cite{Zaslavsky2000} and later by Venegeroles \cite{Venegeroles2008}. A process of \textit{type II} leads to diffusion into islands of the standard map \cite{PhysRevE.85.066210}.

In the present work suppression of accelerator modes by randomness of \textit{type I} and \textit{type II} will be studied. The outline of this paper is as follows. In \Sec{ensembles} approximate analytical expressions for the diffusion coefficient in phase space of the standard map will be derived for processes of \textit{type I} and \textit{type II}.
These will be compared and tested numerically in \Sec{Results}. the results will be summarized and discussed in \Sec{Summary}

\section{Ensembles of standard maps}\label{sec:ensembles}
The standard map is defined by \Eq{Standard map}. For any $K>0$  chaotic regions in phase space are found. There is a critical value $K=K_c\simeq0.9716$ so that for $K<K_c$ the various chaotic regions are separated by invariant circles while for $K>K_c$ chaotic regions merge so that there is an infinite chaotic component and diffusion in momentum is found in numerical calculations for many values of $K$. 

The diffusion coefficient was calculated as an expansion series in powers of $\tfrac{1}{\sqrt{K}}$ \cite{Rechester1981}. To define and calculate the diffusion coefficient, noise was introduced by the distribution of the random variable $\delta\theta$ \cite{Karney1982425,Rechester1980,Rechester1981},

\begin{equation}
\eta(\delta\theta,J)=\frac{1}{\sqrt{2\pi}\sigma}\sum_{n=-\infty}^{\infty}e^{-\frac{\left(\delta\theta-J+2\pi n\right)^{2}}{2\sigma^{2}}}
\end{equation}
which centers $\delta\theta$ around a mean value equal to $J+2\pi n$ as in \Eq{Standard map}. It may be replaced with help of the Poisson summation formula by
\begin{equation}
\eta(\delta\theta,J)=\frac{1}{\sqrt{2\pi}\sigma}\sum_{m=-\infty}^{\infty}e^{-\frac{1}{2}\sigma^{2}m^{2}+im(\delta\theta-J)}.\label{eq:3}
\end{equation}
The deterministic evolution \Eq{Standard map} in time is then
replaced by a probabilistic one given by the distribution
\begin{equation}
P\left(\theta,J,t\right)=\intop^{2\pi}_{0} d\theta'\eta\left(\theta-\theta',J\right)P\left(\theta',J+K\
sin\left(\theta'\right),t-1\right)\label{eq:Chapman}
\end{equation}
where $\theta,J$ are the values at time $t$ and $\theta',J+K\sin\theta'$
are the values at time $t-1$. Considering the initial probability density $P(\theta,J,t=0)=\frac{1}{2\pi}\delta(J-J_0)$ the diffusion coefficient in momentum
space to leading order in $\tfrac{1}{\sqrt{K}}$ is

\begin{equation}
D_{SM}(K)\approx(\frac{K}{2})^{2}\left[1-2\mathcal{J}_{2}(K)e^{-\sigma^{2}}\right]\label{eq:DSM}
\end{equation}
where $\mathcal{J}_n(K)$ are the Bessel functions of order $n$, as found by RRW \cite{Rechester1981}. For large $K$ it reduces approximately to 
\begin{equation}
D_{QL}:=\frac{K^{2}}{4}.
\end{equation}
In the limit $\sigma\to0$, Eq.\Eq{DSM} approximates extremely
well the momentum diffusion coefficient for nearly all $K>K_{c}$,
excluding small intervals near $K\thickapprox2\pi n$ (with integer
$n$) where the acceleration modes are found. Although the limit  $\sigma\to0$ of \Eq{DSM} exists, for the standard map \Eq{Standard map} without noise the diffusion coefficient is not defined, as result of ``sticking'' to regular structures. In the present work ensembles of standard maps will be introduced in terms of the distribution of the parameter $K$. As explained in the introduction it describes physical situations where either the value of the parameter $K$ is not known exactly but is fixed or it is not known and it is varying. We will assume that the parameter $K$ is taken from a Gaussian distribution with average $\bar{K}$ and standard deviation $\sigma_K$ 
\begin{equation}
P(K)=\frac{1}{\sqrt{2\pi}\sigma_K}e^{-\frac{\left(K-\bar{K}\right)^{2}}{2\sigma_K^{2}}}.\label{eq:K distribution}
\end{equation}
This randomness is applied in two ways, defining two types of processes
and ensembles.

\subsection{\textit{Type I}, The parameter $K$ is fixed but not known}\label{sec:Type1}
In this type of dynamics each system is evolved with the parameter
$K$ taking the \textit{same value}, that is chosen at random
from the Gaussian distribution \Eq{K distribution}. The diffusion
coefficient for each system is given by \Eq{DSM} and the average
diffusion coefficient is found by 
\begin{equation}
D_{F}(\bar{K},\sigma,\sigma_K)=\intop_{-\infty}^{\infty} dKP\left(K\right)D_{SM}\left(K\right)\label{eq:DFA-1}.
\end{equation}
In the leading order in $\sigma_{K}$ it reduces to
\begin{equation}
D'_{QL}:=(\frac{\bar{K}}{2})^{2}+\frac{\sigma_K^{2}}{4}.\label{eq:DQL-1}
\end{equation}
The symbol $F$ stands for averaging over final results of $D_{SM}(K)$.
\subsection{\textit{Type II}, Changing parameter at each step}\label{sec:type2}
In this type each system is evolved in a way that $K$
changes at random at each step and is chosen at random with the distribution \Eq{K distribution}.
In this case the distribution \Eq{Chapman} is replaced by
\begin{equation}
P\left(\theta',J+K\sin\left(\theta'\right),t-1\right)\rightarrow\left\langle P\left(\theta',J+K\sin\left(\theta'\right),t-1\right)\right\rangle _{K}
\end{equation}
where $\left<\right>_{K}$ denotes averaging with respect to \Eq{K distribution}, leading to 
\begin{equation}
P\left(\theta,J,t\right)=\intop^{2\pi}_{0} d\theta'\eta\left(\theta-\theta',J\right)\intop_{-\infty}^{\infty} dKP(K)P\left(\theta',J+K\sin\left(\theta'\right),t-1\right).\label{eq:chapman with average}
\end{equation}
Taking the initial probability 
\begin{equation}
P\left(\theta,J,0\right)=\frac{1}{2\pi}\delta(J-J_{0})
\end{equation}
and calculating the paths in Fourier space following \cite{Rechester1981} one obtains an expression in next to leading order in $\tfrac{1}{\sqrt{K}}$  for the diffusion coefficient: 
\begin{equation}
D_{E}(\bar{K},\sigma,\sigma_K)\approx(\frac{\bar{\ensuremath{K}}}{2})^{2}+\frac{\sigma_K^{2}}{4}-2(\frac{\bar{\ensuremath{K}}}{2})^{2}\sum_{l'=-\infty}^{\infty}\mathcal{J}_{2l'+2}(\bar{\ensuremath{K}})\cdot \mathcal{I}_{l'}(\frac{\sigma_K^{2}}{4})\cdot e^{-\sigma^{2}}\cdot e^{-\frac{\sigma_K^{2}}{4}}\label{eq:DES}
\end{equation}
where $\mathcal{J}_{l}$ and $\mathcal{I}_{l}$ are the first order Bessel functions and
first order modified Bessel functions respectively. $E$ stands for
changing parameter at Each step. The details of this calculation are presented
in Appendix A. In the leading order in $\tfrac{1}{\sqrt{K}}$ where
the sum can be neglected, and one obtains for \Eq{DES}  the same result as \Eq{DQL-1}. That is, in the leading order in $\tfrac{1}{\sqrt{K}}$, $D_F=D_E$

\section{Comparison of the results found for the two different ensembles}\label{sec:Results}
\InsertFigTwo{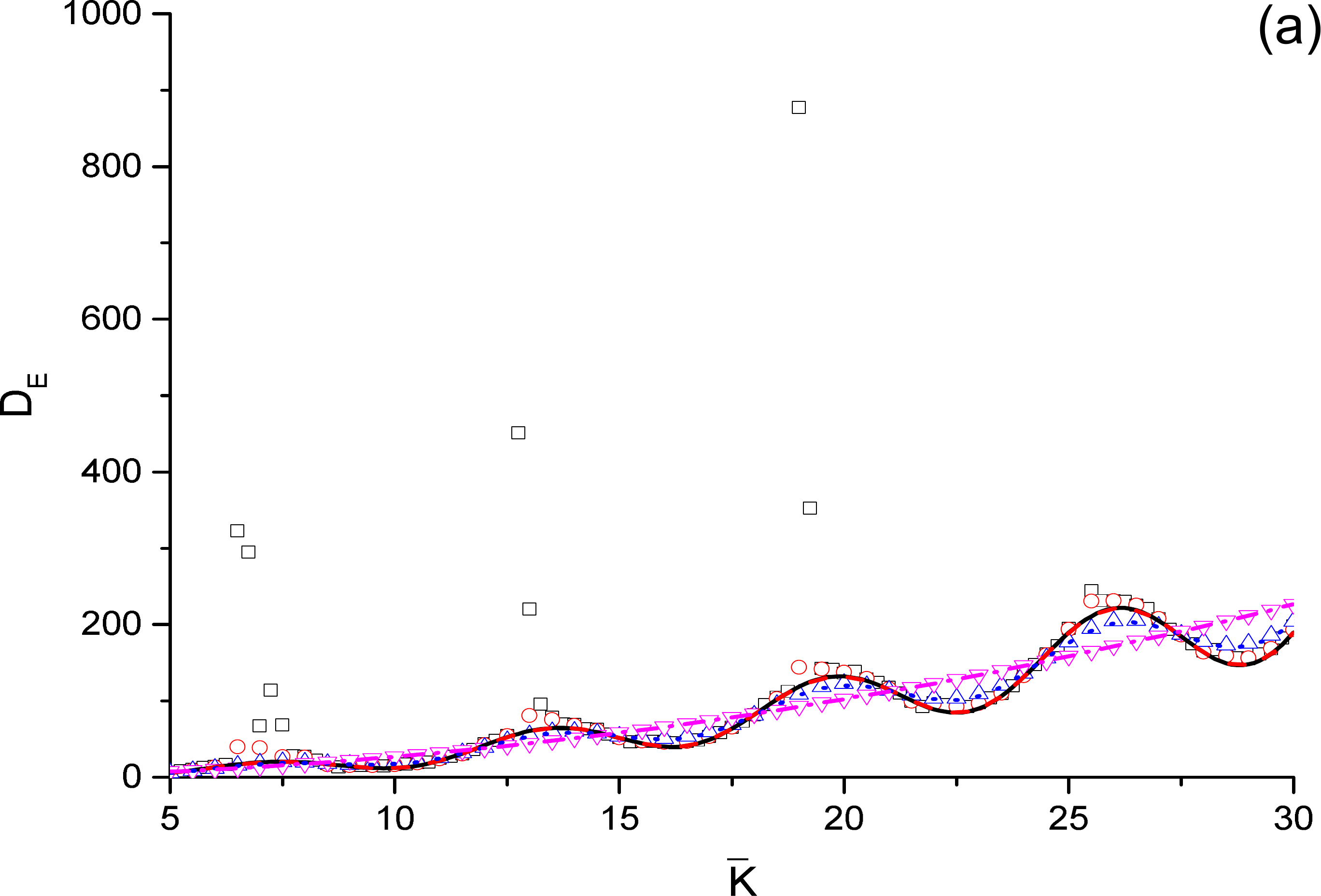}{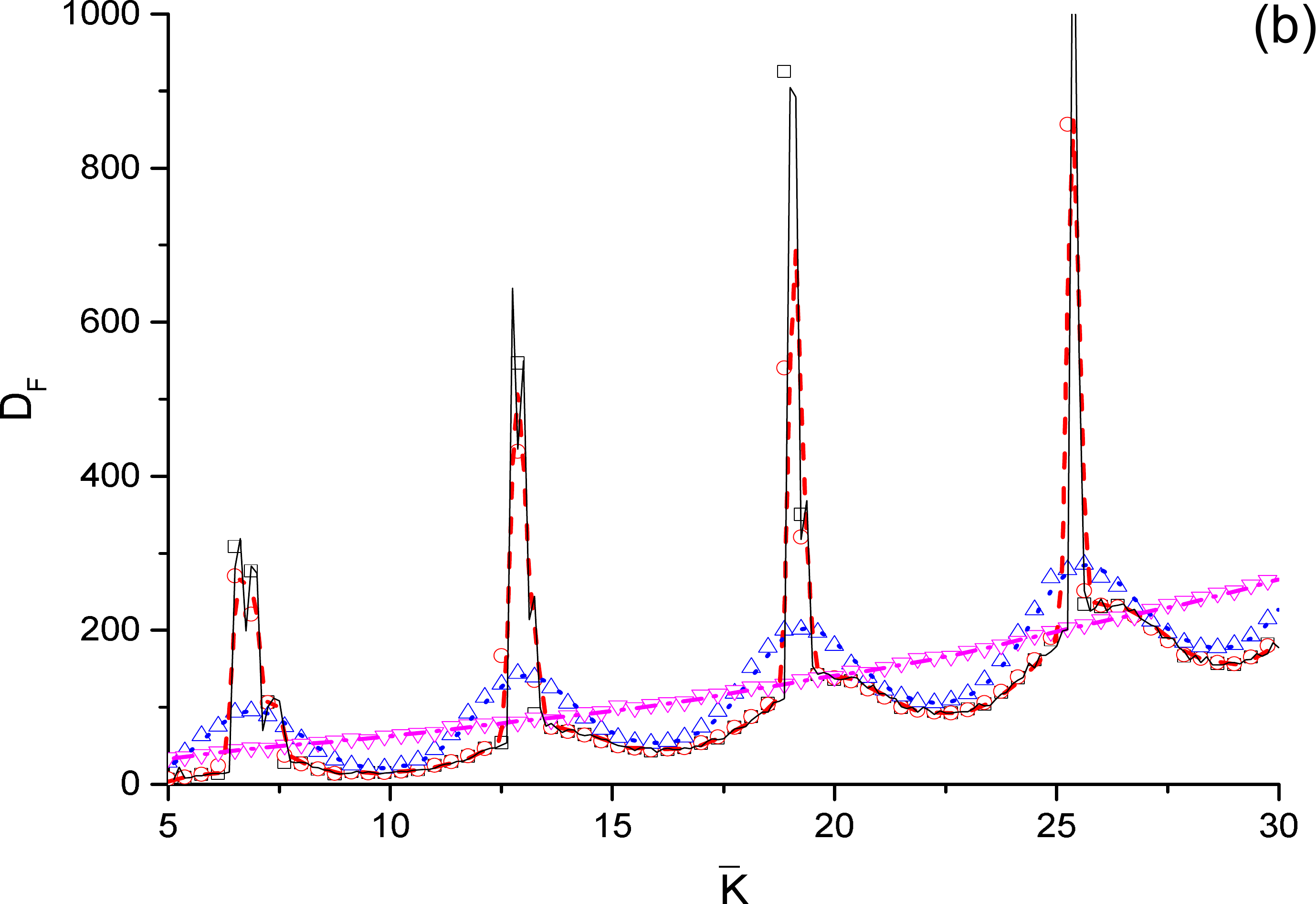}{(Color online) Diffusion coefficients found by numerical simulations (analytic results of \Eq{DES} and \Eq{DFA-1}) (a) $D_E$ and (b) $D_F$ for problems of \textit{type II} and \textit{I} respectively, as a function of $\bar{K}$ for various values of $\sigma_K$:  $\sigma_K=0.1$  in red  $\circ$ ($- -$), $\sigma_K=1$ in blue $\triangle$ ($\dotsb$), $\sigma_K=3$ in magenta $\triangledown$ ($-.-$), and $\sigma_K=1\times10^{-5}$ black $\square$  ($-$). The black line in (b) was calculated from \Eq{Diffusion} contrary to other cases. In the calculation of $D_F$ by \Eq{DFA-1}, $D_{SM}$ was obtained from a numerical simulation.}{Fig1}{0.55}
In this section the values of the diffusion coefficient $D_{F}$ \Eq{DFA-1}
and $D_{E}$ \Eq{DES} will be compared, and comparison to the results of numerical simulations will be presented. For both processes, $I$ and $II$, $N=10,000$ initial conditions were used for each value of $K$ and the map was
iterated $N_t=100$ steps . The number of values of $K$ chosen from the
ensemble defined by \Eq{K distribution} is $N'=10,000$. The map
\Eq{Standard map} was evolved in the presence of noise with
the distribution \Eq{3}. The variance $\sigma$ of the noise was set to a small value, namely $10^{-5}$ and was kept constant, while the variance of $K$,
$\sigma_K$, was varied and the diffusion coefficient was calculated by \Eq{DFA-1} and \Eq{DES}. The corresponding numerical calculations were done for 200 different values of $\bar{K}$ in the interval  $\bar{K}_{min}=5$ and $\bar{K}_{max}=30$. For a given $K$ the diffusion coefficient was calculated numerically by 
\begin{equation}\label{eq:Numeric}
D(\bar{K},\sigma,\sigma_K)=\tfrac{\left<(J-\left<J\right>)^2\right>}{2N_t}.
\end{equation}

\Fig{Fig1} presents the numerical results of $D_{E}$ and $D_{F}$ for various values of $\sigma_K$. It is shown that while the oscillations in $D_{E}$ decay with increasing the value of $\sigma_K$, the  values of $D_{F}$ exhibit enhanced diffusion in the entire range compared to $D_E$. This is due to the effect of convolution of the Gaussian probability with the accelerator modes. Furthermore, the accelerator modes disappear in $D_{E}$ for smaller value of $\sigma_K$ than they do for $D_{F}$, Therefore $D_F$ is time dependent.
 
In \Fig{analyticsnumerics} the diffusion coefficient is presented for moderate $\sigma_K$=1. Numerical and analytical calculations are shown for the two types and compared with \Eq{DSM} where $K$ is replaced by $\bar{K}$. Depletion or blurring of the accelerator modes is clearly shown for  \textit{type II} processes, such that for this value of $\sigma_K$ the diffusion is well described by \Eq{DES}. RRW \cite{Rechester1981} found that the peak of the accelerator mode is obtained by summation Fourier paths of very high order (Fig 9 in \cite{Rechester1981}). Since there is no exact analytic expression for the accelerator modes, we found numerically $D(K)$ of \Eq{Diffusion} with only RRW type of noise, and used it in \Eq{DFA-1}. For $\sigma_K= 10^{-5}$ we assumed $D_F(\bar{K})=D(\bar{K})$. The time dependence of $D_F$ (resulting from the accelerator modes) is demonstrated in the inset of \Fig{analyticsnumerics}
  
\InsertFig{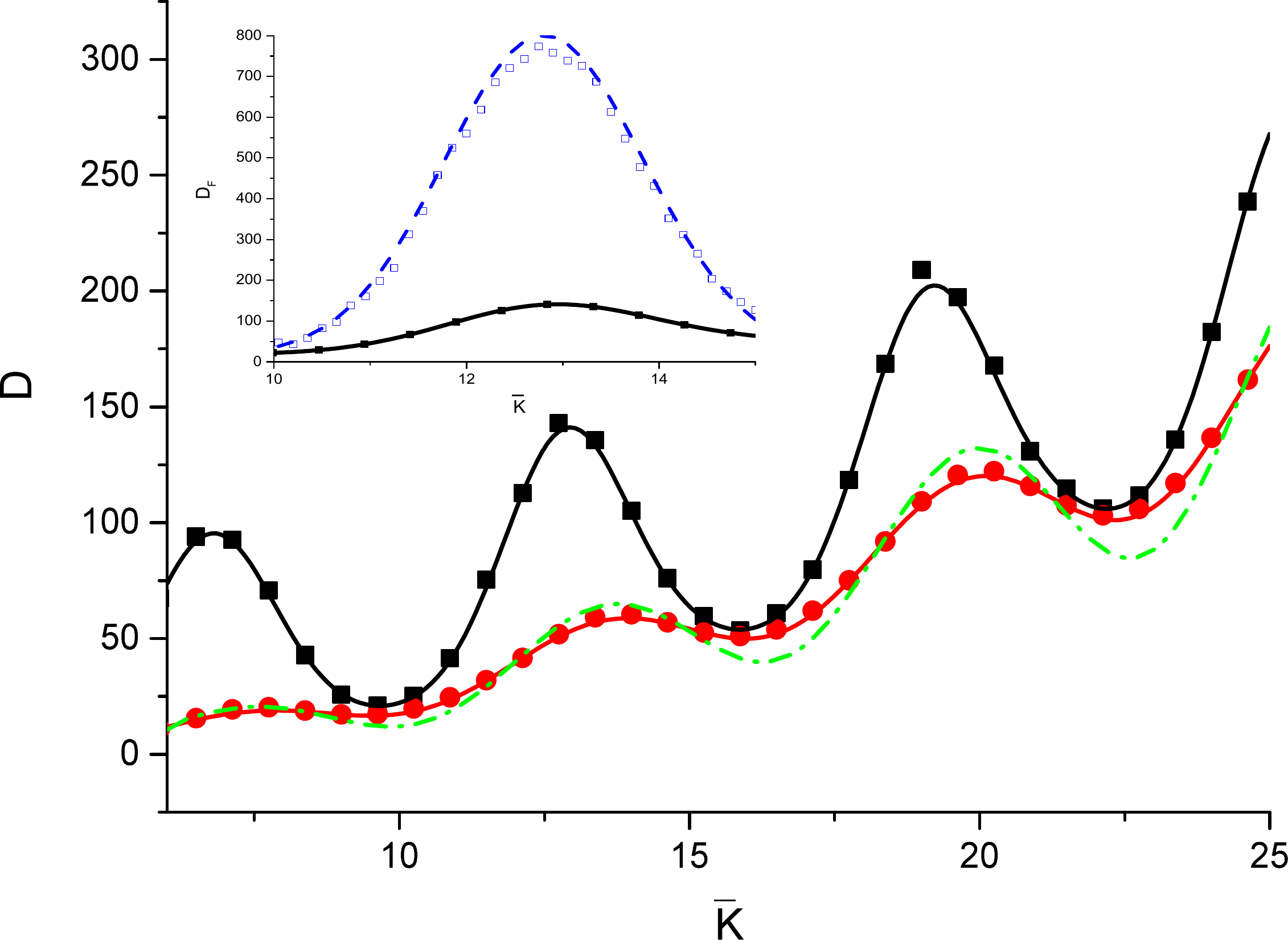}{(Color online) Comparison between analytical and numerical calculations of $D_F$ and $D_E$ for $\sigma_K=1$. Numerical results for $D_F$ shown in black $\blacksquare$, and for $D_E$ are in red \textcolor{red}{$\bullet$}. Analytic calculation for $D_F$ \Eq{DFA-1} is represented by black solid line, and for $D_E$ \Eq{DES} in gray (red). The analytical results of $D_{SM}$ \Eq{DSM} is represented by dashed dotted (green) line. The inset shows the numerical results for $D_F$ for $N_t=1000$ in blue \textcolor{blue}{$\square$} , and for $N_t=100$ in black $\blacksquare$. The analytical results are in blue dashed line and in black solid line correspondingly  }{analyticsnumerics}{0.8}

The elimination of the accelerator modes for the  \textit{type II} system can be explained by the following. Because contribution to it comes from Fourier paths of high order, multiplying each step of the path by $e^{-k'^2\sigma_K^2/4}$ in \Eq{recurssion} affects the contribution strongly. In this sense this type of randomness eliminates the accelerator modes more effectively than the RRW type of noise. Increasing time (not shown here) leads to even further diminishing effect of the accelerator modes. Although longer paths give a contribution, it is negligible due to devision by $1/2N_t$.

\section{Summary}\label{sec:Summary}

In the present work two ensembles of standard maps of the form \Eq{Standard map} with a Gaussian
distribution of the driving parameter $K$ were studied. In one case
(\textit{type I}) the $K$ was kept fixed and finally average over the diffusion
coefficient was taken while in the other process (\textit{type II}) $K$ was
varied at each step. Noise with standard deviation $\sigma= 10^{-5}$ was added to
make the diffusion well defined. The resulting diffusion coefficients
differ as will be discussed in what follows. Increasing $\sigma_K^2$, the variance
of $K$, tends to wash out details, hence the effect of acceleration
modes weakens, as can be seen from Figs. \ref{fig:Fig1}, and
\ref{fig:analyticsnumerics}. We see that relatively weak randomness is
sufficient to wash out the fine details of the dynamics generated
by the map. The most important is the effect of this on accelerator modes. We find that the averaged diffusion coefficient exhibits oscillations as a function of the averaged driving parameter $\bar{K}$ similar to the situation found for fixed $K$ \cite{Rechester1981}. The processes $I$ result in replacement of the acceleration modes by enhanced diffusion for a wide range of $\bar{K}$.  
The process \textit{II} where the driving parameter $K$ is chosen at each step is more effective in washing out the accelerator modes than processes $I$ where $K$ is chosen at random but is kept fixed for the entire evolution time. This leads us to conjecture that processes of the \textit{type II} are more efficient in eliminating fine details of mixed systems.

The processes of types $I$ and $II$ are paradigms of various physical situations and the difference found here may give hope
to develop statistical theories for mixed systems as proposed by \cite{Ceder2013,Meiss1985,Cristadoro2008}
and others, and to classify relevant averaging processes into few classes.

\begin{acknowledgments}
The work was supported in part by the Israel Science Foundation (ISF) grant number
1028/12, and by the US-Israel Binational Science Foundation (BSF) grant
number 2010132, by the Shlomo Kaplansky academic chair.
We would like to thank James Meiss, Edward Ott and Hagar Veksler for fruitful discussions.
\end{acknowledgments}

\appendix

\section{Calculating the diffusion coefficient $D_{E}$}

We start the calculation from (\ref{eq:chapman with average}). The Fourier expansion
is :
\begin{equation}
P\left(\theta,J,t\right)=\frac{1}{(2\pi)^{2}}\sum_{m=-\infty}^{\infty}\intop_{-\infty}^{\infty}dk\, a_{m}^{t}\left(k\right)e^{i(m\theta+kJ)}
\end{equation}

We now turn to find the recursion relation of the expansion coefficient
$a_{m}^{t}\left(k\right)$ substituting the Fourier expansion for
the distribution at time $t-1$, one finds{\footnotesize 
	\begin{equation}
	P\left(\theta,J,t\right)=\intop_{0}^{2\pi}d\theta'\frac{1}{2\pi}\sum_{\tilde{m}=-\infty}^{\infty}e^{-\frac{\sigma^{2}\tilde{m}^{2}}{2}+i\tilde{m}(\theta-\theta'-J)}\frac{1}{(2\pi)^{2}}\sum_{m'=-\infty}^{\infty}\intop_{-\infty}^{\infty}dk\, a_{m'}^{t-1}\intop_{-\infty}^{\infty}dK[\frac{1}{\sqrt{2\pi}\sigma_K}e^{-\frac{(K-\bar{\ensuremath{K}})^{2}}{2\sigma^{'2}}}e^{i(m'\theta'+kJ+kK\sin\theta')}]
	\end{equation}
}{\footnotesize \par}

Integrating over the noise and averaging over $K$ one finds{\footnotesize 
	\begin{multline}
	P\left(\theta,J,t\right)=\\
	\sum_{\tilde{m}=-\infty}^{\infty}e^{-\frac{\sigma^{2}\tilde{m}^{2}}{2}+i\tilde{m}\theta}\sum_{m'=-\infty}^{\infty}\intop_{-\infty}^{\infty}dk'\, a_{m'}^{t-1}(k')e^{-k'^{2}\frac{\sigma_K}{4}}\sum_{l=-\infty}^{\infty}\mathcal{J}_{l}(|k|\bar{\ensuremath{K}})e^{i(k'-\tilde{m)}J}\sum_{l'=-\infty}^{\infty}I_{l'}(\frac{k^{2}\sigma_K^{2}}{4})\delta(m'-\tilde{m}+lsignk'+2l')
	\end{multline}
}Taking the sum over $m'$ and making use of the orthogonality of
the Fourier components one finds the recursion relation of the Fourier
components

\begin{equation}{\label{eq:recurssion}}
a_{m}^{t}(k)=e^{-\frac{\sigma^{2}\tilde{m}^{2}}{2}}e^{-k'^{2}\frac{\sigma_K^{2}}{4}}\sum_{l=-\infty}^{\infty}\mathcal{J}_{l}(|k|\bar{\ensuremath{K}})\sum_{l'=-\infty}^{\infty}\mathcal{I}_{l'}(\frac{k^{2}\sigma_K^{2}}{4})a_{\tilde{m}-lsignk'-2l'}^{t-1}(k')
\end{equation}
with the following relations :
\begin{equation}
\begin{array}{c}
\tilde{m}=m\\
k'=k+m\\
m'=m-lsignk'-2l'
\end{array}\label{eq:relations}
\end{equation}
Next using paths in Fourier space the leading order in the Bessel
functions can evaluated. The Diffusion coefficient is 
\begin{equation}
D=\lim_{t\rightarrow\infty}\frac{\left\langle (J_{t}-J_{0})^{2}\right\rangle }{2t}\label{eq:Diffusion Coef}
\end{equation}
where \cite{Rechester1981} 
\begin{equation}
\left\langle (J_{t}-J_{0})^{2}\right\rangle =i^{2}\lim_{k\rightarrow0^{+}}\left(\frac{\partial^{2}}{\partial k^{2}}\right)a_{0}^{t}(k).\label{eq:moment generating}
\end{equation}
The leading order term is the path which stays at the origin for t
steps that is 
\begin{equation}
a_{0}^{t}(k)=\left[e^{-k'^{2}\frac{\sigma_K^{2}}{4}}\sum_{l=-\infty}^{\infty}\mathcal{J}_{l}(|k|\bar{\ensuremath{K}})\sum_{l'=-\infty}^{\infty}\mathcal{I}_{l'}(\frac{k^{2}\sigma_K^{2}}{4})\right]^{t}a_{0}^{0}(0)
\end{equation}
since when differentiating twice and taking $k\rightarrow0^{+}$ limit
only the $J_{0},I_{0}$ contribute, one finds
\begin{equation}
a_{0}^{t}\simeq\left[e^{-k^{2}\frac{\sigma_K}{4}}\left[1-\left(\frac{k\bar{\ensuremath{K}}}{2}\right)^{2}\right]\left[1+\left(\frac{k^{2}\sigma_K^{2}}{8}\right)^{2}\right]\right]^{t}.
\end{equation}
The leading term is (\ref{eq:DQL-1})

The next term is calculated using a path that leaves the origin. leaving
and returning to the origin has to be done using two different points.
The shortest path possible is of the form
\begin{equation}
0,0\rightarrow k,m\rightarrow k',m'\rightarrow0,0
\end{equation}
using the relations (\ref{eq:relations}), when leaving the origin
\begin{equation}
-m=-l-2l'
\end{equation}
and when entering the origin 
\begin{equation}
m'=0-l-2l'.
\end{equation}
The only paths making a contribution to the diffusion are like in
\cite{Rechester1980,Rechester1981} 
\begin{equation}
\begin{array}{c}
(0,0)\rightarrow(1,-1)\rightarrow(0,1)\rightarrow(0,0)\\
(0,0)\rightarrow(-1,1)\rightarrow(0,-1)\rightarrow(0,0)
\end{array}
\end{equation}
and each has an equal contribution. For the second transition in the
first path we get 
\begin{equation}
1=-1+l-2l'
\end{equation}
or
\begin{equation}
l=2+2l'
\end{equation}
leading to 
\begin{equation}
\begin{array}{c}
a_{0}^{t}(k)=2(t-2)\left(\sum_{l'=-\infty}^{\infty}\mathcal{J}_{2l'}(|k|\bar{\ensuremath{K}})\mathcal{I}_{l'}(\frac{k^{2}\sigma_K^{2}}{4})\right)^{t-3}\sum_{l'=-\infty}^{\infty}\mathcal{J}_{-2l'-1}(|k|\bar{\ensuremath{K}})\mathcal{I}_{l'}(\frac{k^{2}\sigma_K^{2}}{4})\cdot\\
\cdot\sum_{l'=-\infty}^{\infty}\mathcal{J}_{2l'+2}(|-1+k|\bar{\ensuremath{K}})\mathcal{I}_{l'}(\frac{(-1+k)^{2}\sigma_K^{2}}{4})e^{-\frac{\sigma^{2}}{2}}e^{-(-1+k)^{2}\frac{\sigma_K^{2}}{4}}\sum_{l'=-\infty}^{\infty}\mathcal{J}_{-2l'-1}(|k|\bar{\ensuremath{K}})\mathcal{I}_{l'}(\frac{k^{2}\sigma_K^{2}}{4})e^{-\frac{\sigma}{2}}
\end{array}
\end{equation}
using (\ref{eq:Diffusion Coef}) and (\ref{eq:moment generating})
and including the leading term one finds (\ref{eq:DES}).

\bibliographystyle{unsrtnat}
\bibliography{Diffusion2}

\end{document}